\documentclass[11pt,twoside,a4paper,cmspaper,final,collab]{cms-tdr}
\def\svnVersion{21ea23c}\def\svnDate{2026/06/27}\def\cmsCernNoTag{CERN-EP-2025-273}\def\cmsCernDate{\today}\def\cmsMessage{Published in the Journal of High Energy Physics as \href{http://dx.doi.org/10.1007/JHEP06(2026)187}{\doi{10.1007/JHEP06(2026)187}.}}
\begin{document}\cmsNoteHeader{SMP-24-019}

\newcommand{\emu}{\ensuremath{\Pe\PGm}\xspace}
\newcommand{\mumu}{\ensuremath{\PGm\PGm}\xspace}
\providecommand{\PGgst}{{\HepParticle{\PGg}{}{\ast}}\xspace}
\newcommand{\ptemu}{\ensuremath{\pt^{\Pe\PGm}}\xspace}
\newcommand{\ggtotautau}{{\ensuremath{\PGg\PGg\to\PGt\PGt}}\xspace}
\newcommand{\ggtomumu}{{\ensuremath{\PGg\PGg\to\PGm\PGm}}\xspace}
\newcommand{\ggtoee}{{\ensuremath{\PGg\PGg\to\Pe\Pe}}\xspace}
\newcommand{\ggtoww}{{\ensuremath{\PGg\PGg\to\PWp\PWm}}\xspace}
\newcommand{\ntracks}{\ensuremath{N_\text{tracks}}\xspace}
\newcommand{\ntracksHS}{\ensuremath{N_\text{tracks}^\text{HS}}\xspace}
\newcommand{\ntracksPU}{\ensuremath{N_\text{tracks}^\text{PU}}\xspace}
\newcommand{\Wjets}{\ensuremath{\PW+\text{jets}}\xspace}
\newcommand{\Zjets}{\ensuremath{\PZ/\PGgst+\text{jets}}\xspace}
\newcommand{\mmumu}{\ensuremath{m_{\PGm\PGm}}\xspace}
\newcommand{\mZ}{\ensuremath{m_{\PZ}}\xspace}
\newcommand{\mW}{\ensuremath{m_{\PZ}}\xspace}
\newlength\cmsFigWidth\setlength\cmsFigWidth{0.45\textwidth}
\newlength\cmsTabSkip\setlength{\cmsTabSkip}{1ex}

\cmsNoteHeader{SMP-24-019}
\title{Measurement and effective field theory interpretation of the photon-fusion production cross section of a pair of \texorpdfstring{\PW}{W} bosons in proton-proton collisions at \texorpdfstring{$\sqrt{s}=13\TeV$}{sqrt(s) = 13 TeV}}

\date{\today}

\abstract{This analysis presents an observation of the photon-fusion production of $\PW$ boson pairs using the CMS detector at the LHC. The total cross section of the $\PW^+\PW^-$ production in photon fusion is measured using proton-proton collision data with an integrated luminosity of 138\fbinv collected with the CMS detector in 2016--2018 at a center-of-mass energy of $\sqrt{s}=13\TeV$. Events are selected in the final state with one isolated electron and one isolated muon, and no additional tracks associated with the electron-muon production vertex. The total and fiducial production cross sections are $643^{+82}_{-78}\unit{fb}$ and $3.96^{+0.53}_{-0.51}\unit{fb}$, respectively, in agreement with the standard model predictions of $631\pm 126\unit{fb}$ and $3.87\pm 0.77\unit{fb}$. This agreement enables stringent constraints on anomalous quartic gauge couplings within a dimension-8 effective field theory framework.  
}

\hypersetup{%
pdfauthor={CMS Collaboration},%
pdftitle={Measurement and effective field theory interpretation of the photon-fusion production cross section of a pair of W bosons in proton-proton collisions at sqrt(s) = 13 TeV},%
pdfsubject={CMS},%
pdfkeywords={CMS, W, photon, SM, EFT}}

\maketitle 

\section{Introduction}

Photon-induced processes in proton-proton ($\Pp\Pp$) collisions at the LHC provide important tests of the standard model (SM) of particle physics, which are complementary to those accessible through the more common quark and gluon interactions. In the SM, the photon-fusion production of a pair of $\PW$ bosons proceeds through interactions that involve trilinear and quartic gauge boson couplings.
The measurement of this process can be used to constrain dimension-8 effective field theory (EFT) operators related to anomalous quartic gauge boson interactions.

{\tolerance=2400
The signal events considered in this analysis correspond to the process $\Pp\Pp (\PGg\PGg) \to \Pp^{(*)} \PWp \PWm \Pp^{(*)}$, where  $\Pp^{(*)}$
denotes a final-state proton that either stays intact or dissociates after the photon emission. When both protons coherently radiate a photon and both remain intact after their interaction, the process is called exclusive or elastic.
When one or both incoming protons fragment after the photon emission, the process is either semi- or fully dissociative. Semi- and fully dissociative events can also give rise to an elastic-like signature when the proton remnants are not
reconstructed because they fall outside of the detector acceptance.
\par}

The selection of signal events requires the absence of additional charged particles around the production vertex, as is characteristic of photon-fusion 
processes. 
Such a strategy was previously used by the CMS Collaboration to observe the $\ggtotautau$ process
in $\Pp\Pp$ collisions~\cite{CMS:2024qjo}
and in earlier searches and measurements at the LHC~\cite{CMS:2011vma,CMS:2013hdf,CMS:2016rtz,CMS:2018uvs,ATLAS:2016lse,ATLAS:2020iwi}.
Evidence for the photon-fusion production of a pair of $\PW$ bosons was reported by the CMS Collaboration using 
data collected at center-of-mass energies of 7 and 8\TeV~\cite{CMS:2016rtz}, and this process was then observed by the ATLAS Collaboration 
with data collected at a center-of-mass energy of 13\TeV~\cite{ATLAS:2020iwi}. Limits were previously imposed on CP-even dimension-8 EFT operators using the $\ggtoww$ process by both the ATLAS~\cite{ATLAS:2016lse} and CMS~\cite{CMS:2016rtz} Collaborations.

The present analysis relies on $\Pp\Pp$ collision data collected at a center-of-mass energy of 13\TeV with the CMS detector in 2016--2018, 
corresponding to an integrated luminosity of 138\fbinv.  
 The $\ggtoww$ process is measured in the final
state with an isolated electron, an isolated muon of opposite charge, and no additional tracks associated with the vertex formed by the electron and the muon. 
The dominant backgrounds are the inclusive production of \PW boson pairs from all sources except the signal $\ggtoww$, as well as $\PZ/\PGgst\to\PGt\PGt$ events, which are constrained from 
data in selected events with a larger number of tracks. 
The cross section of the $\ggtoww$ process is measured both inclusively and in a fiducial phase space matching the analysis selection criteria. 
Potential beyond-the-SM (BSM) contributions are investigated using an EFT extension of the SM, from which stringent constraints on dimension-8 operators are derived, including for the first time limits on CP-odd dimension-8 operators.   

The paper is structured as follows. Sections~\ref{sec:detector} and~\ref{sec:simulations} describe the detector and the simulation of signal and background processes. The object reconstruction is detailed in
Section~\ref{sec:object}, followed by the event selection in Section~\ref{sec:selection}. The analysis strategy and methods to estimate signal and background processes are presented in Sections~\ref{sec:strategy} and~\ref{sec:background}. 
After a discussion of the systematic uncertainties in Section~\ref{sec:systematics}, the results and the EFT interpretation are reported in Section~\ref{sec:results}.
The paper is summarized in Section~\ref{sec:summary}. Tabulated results are provided in the HEPData record for this analysis~\cite{hepdata.167352}.

\section{The CMS detector}\label{sec:detector}

The CMS apparatus~\cite{CMS:2008xjf,CMS:2023gfb} is a multipurpose, nearly hermetic detector, designed to trigger on~\cite{CMS:2020cmk,CMS:2016ngn,CMS:2024aqx} and identify electrons, muons, photons, and (charged and neutral) hadrons~\cite{CMS:2020uim,CMS:2018rym,CMS:2014pgm}. Its central feature is a superconducting solenoid of 6\unit{m} internal diameter, providing a magnetic field of 3.8\unit{T}. Within the solenoid volume are a silicon pixel and strip tracker, a lead tungstate crystal electromagnetic calorimeter (ECAL), and a brass and scintillator hadron calorimeter (HCAL), each composed of a barrel and two endcap sections. Forward calorimeters extend the pseudorapidity ($\eta$) coverage provided by the barrel and endcap detectors. Muons are reconstructed using gas-ionization detectors embedded in the steel flux-return yoke outside the solenoid. More detailed descriptions of the CMS detector, together with a definition of the coordinate system used and the relevant kinematic variables, can be found in Refs.~\cite{CMS:2008xjf,CMS:2023gfb}.

Events of interest are selected using a two-tiered trigger system. The first level (L1), composed of custom hardware processors, uses information from the calorimeters and muon detectors to select events at a rate of around 100\unit{kHz} within a fixed latency of 4\mus~\cite{CMS:2020cmk}. The second level, known as the high-level trigger (HLT), consists of a farm of processors running a version of the full event reconstruction software optimized for fast processing, and reduces the event rate to a few kHz before data storage~\cite{CMS:2016ngn,CMS:2024aqx}.

\section{Signal and background processes}\label{sec:simulations}

Monte Carlo simulation samples corresponding to the elastic, semi-dissociative, and fully dissociative photon-fusion productions of pairs of \PGt leptons and \PW bosons are generated with the \textsc{SuperChic}~4 generator~\cite{Harland-Lang:2020veo,Bailey:2022wqy}. 
Elastic, semi-dissociative, and dissociative components are simulated independently and combined together according to the predictions 
from the \textsc{SuperChic}~4 generator. Semi-dissociative and dissociative components amount to about five times the elastic contribution, for the selection criteria used in the analysis and detailed in Section ~\ref{sec:selection}.  The $\ggtoww$ processes in which the observed electron or muon originates from a leptonic \PGt decay, with the \PGt lepton itself produced in a \PW boson decay, are also simulated and 
treated as a component of the signal. 

The \textsc{SuperChic} generator is used to model the SM prediction of the signal, and the \textsc{gamma-UPC} generator~\cite{Shao:2022cly} is used to predict relative variations from nonzero dimension-8 EFT operators. 
The \textsc{gamma-UPC} generator,  which is implemented in the \MGvATNLO v3.5.6 code~\cite{Alwall:2014hca,Frederix:2012ps}, derives photon fluxes from electric dipole and charge form factors for protons and ions, 
and includes hadronic survival probabilities. In this work, the photon fluxes used are those derived from the elastic charge form factor of the proton, including the effects of the photon virtualities~\cite{Shao:2024dmk}, whereas the \textsc{SuperChic} event generator applies form factors based on the 'structure-function' approach~\cite{Han:1992hr}.
The EFT dimension-8 operators are simulated with \MGvATNLO
at leading order (LO) accuracy using the reweighting technique~\cite{Mattelaer:2016gcx}. We do not include the BSM effects on the background processes since they are expected to have negligible contributions. 

The \MGvATNLO~2.6.5 event generator is used to generate background events originating from the Drell--Yan (DY) dilepton production
process $\Pp\Pp\to\PZ/\PGgst$, as well as the inclusive diboson production, $\Pp\Pp\to\PW\PW$, including the electroweak (EW) $\PW\PW$ production in association with two jets, except for processes involving top quarks, which are simulated separately as described below. These processes are simulated
 at next-to-LO (NLO) accuracy in perturbative quantum chromodynamics (QCD) with up to two outgoing partons at Born level, with different jet multiplicities treated with the FxFx jet matching and merging procedure~\cite{Frederix:2012ps}.
The \MGvATNLO generator is also used for the simulation of the minor contribution from other diboson production events, whereas
\POWHEG 2.0~\cite{Nason:2004rx,Frixione:2007vw, Alioli:2010xd, Alioli:2010xa, Alioli:2008tz,Frixione_2007, Re:2010bp, Frederix:2012dh} is
 used for \ttbar and single top quark production. Contributions from EW $\PW\PZ$ and $\PZ\PZ$ production in association with two jets, as well as double parton scattering $\PW\PW$ are at a percent level and 
 are therefore neglected. 

The generators are interfaced with
 \PYTHIA 8.240~\cite{Sjostrand:2014zea} to model parton showering and hadronization, the underlying event activity, as well as the decay of the
 $\PGt$ leptons. For the signal samples, the modeling of elastic and proton-dissociated contributions follows the procedure described in 
 Ref.~\cite{Harland-Lang:2020veo}. The \PYTHIA generator with {CP5} tune~\cite{CMS:2019csb} is used for the description of the underlying event.
The \textsc{NNPDF3.1} parton distribution function (PDF) set~\cite{Ball:2017nwa} at next-to-NLO accuracy is used for background simulations.
Additional $\Pp\Pp$ interactions per bunch crossing, known as pileup (PU), are modeled with \PYTHIA and added to the simulated samples with a distribution that is chosen to match that observed in the data.
Generated events are processed through a \GEANTfour~\cite{Agostinelli:2002hh} simulation of the CMS detector.

\section{Object reconstruction}\label{sec:object}

A particle-flow algorithm~\cite{CMS:2017yfk} is used to reconstruct and identify each individual particle in an event, with an optimized combination of information from the various elements of the CMS detector. The energy of photons is obtained from the ECAL measurement. The energy of electrons is determined from a combination of the electron momentum at the primary interaction vertex as determined by the tracker, the energy of the corresponding ECAL cluster, and the energy sum of all bremsstrahlung photons spatially compatible with originating from the electron track. The energy of muons is obtained from the curvature of the corresponding track. The energy of charged hadrons is determined from a combination of their momentum measured in the tracker and the matching ECAL and HCAL energy deposits, corrected for the response function of the calorimeters to hadronic showers. Finally, the energy of neutral hadrons is obtained from the corresponding corrected ECAL and HCAL energies. 

Electrons are measured in the pseudorapidity range $\abs{\eta} < 2.5$. The momentum resolution for electrons with transverse momentum $\pt \approx 45\GeV$ from $\PZ \to \Pe \Pe$ decays ranges from 1.6 to 5\%. It is generally better in the barrel region than in the endcaps, and also depends on the bremsstrahlung energy emitted by the electron as it traverses the material in front of the ECAL~\cite{CMS:2020uim,CMS-DP-2020-021}. Electrons are identified with a multivariate discriminant that includes isolation variables, as described in Ref.~\cite{CMS:2020uim}.

Muons are measured in the pseudorapidity range $\abs{\eta} < 2.4$.
The efficiency to reconstruct and identify muons is greater than 96\%. Matching muons to tracks measured in the silicon tracker results in a relative transverse momentum resolution, for muons with \pt up to 100\GeV, of 1\% in the barrel and 3\% in the endcaps. 
The \pt resolution in the barrel is better than 7\% for muons with \pt up to 1\TeV~\cite{CMS:2018rym}. 

The relative isolation of electrons (muons) is calculated using the \pt of tracks in a cone of $\Delta R = \sqrt{\smash[b]{(\Delta \eta)^2 + (\Delta \phi)^2}} < 0.3$ (0.4) centered
on the lepton track, divided by the lepton \pt, corrected for charged and neutral pileup contributions~\cite{CMS:2020uim} (\cite{CMS:2018rym}). A threshold of 0.20 is used, although this requirement is relaxed in certain cases to estimate specific background contributions, as detailed later. 

The silicon tracker used in 2016 measured charged particles within the range $\abs{\eta} < 2.5$. For nonisolated particles of $1 < \pt < 10\GeV$ and $\abs{\eta} < 1.4$, the track resolutions were typically 1.5\% in \pt and 25--90 (45--150)\mum in the transverse (longitudinal) impact parameter~\cite{CMS:2014pgm}. For tracks with \pt as low as 0.5\GeV, the transverse (longitudinal) impact parameter resolution was around 150 (300) \mum. At the start of 2017, a new pixel detector was installed~\cite{Phase1Pixel}; the upgraded tracker measured particles with typical resolutions of 1.5\% in \pt and 20--75\mum in the transverse impact parameter~\cite{DP-2020-049} for nonisolated particles of $1 < \pt < 10\GeV$. The transverse impact parameter resolution was up to about 105\mum for $\pt\simeq0.5\GeV$. 

During the 2016 and 2017 data taking, a gradual shift in the timing of the inputs of the ECAL L1 trigger in the region at $\abs{\eta} > 2.0$ caused a specific trigger inefficiency. For events containing an electron (a jet) with \pt larger than $\approx$50\GeV ($\approx$100\GeV), in the region $2.5 < \abs{\eta} < 3.0$ the efficiency loss is $\approx$10--20\%, depending on \pt, $\eta$, and time. Correction factors were computed from data and applied to the acceptance estimated with the simulation. 

\section{Event selection}\label{sec:selection}

The signal region (SR) and control region (CR) of the analysis feature an $\Pe\PGm$ final state. Events are selected 
by triggering on the presence of an electron, a muon, or both an electron and a muon.
Offline, the leading (trailing) lepton \pt is required to be greater than 24 (15)\GeV. 
The electron (muon) is required to have $\abs{\eta}<2.5$ (2.4), to pass the identification criteria~\cite{CMS:2020uim} (\cite{CMS:2018rym}) 
and be isolated. 
The electron and muon must have opposite-sign (OS) electric charge, be separated by $\Delta R > 0.5$, and have an invariant mass above 20\GeV.
The difference $\abs{d_z}(\Pe,\PGm)$ between the longitudinal impact parameter of the electron and the muon is required to be less than 0.04\cm.
The leptons are selected irrespectively of their $\abs{d_z}$ with respect to the primary vertex.
The $\Pe\PGm$ vertex is then built from the two leptons, and its $z$ position is
calculated as the average $z$ intercepted of the tracks of the two objects.
The event is vetoed if there is any additional identified and isolated electron or muon. 

\begin{figure}
  \centering
  \includegraphics[width=0.45\textwidth]{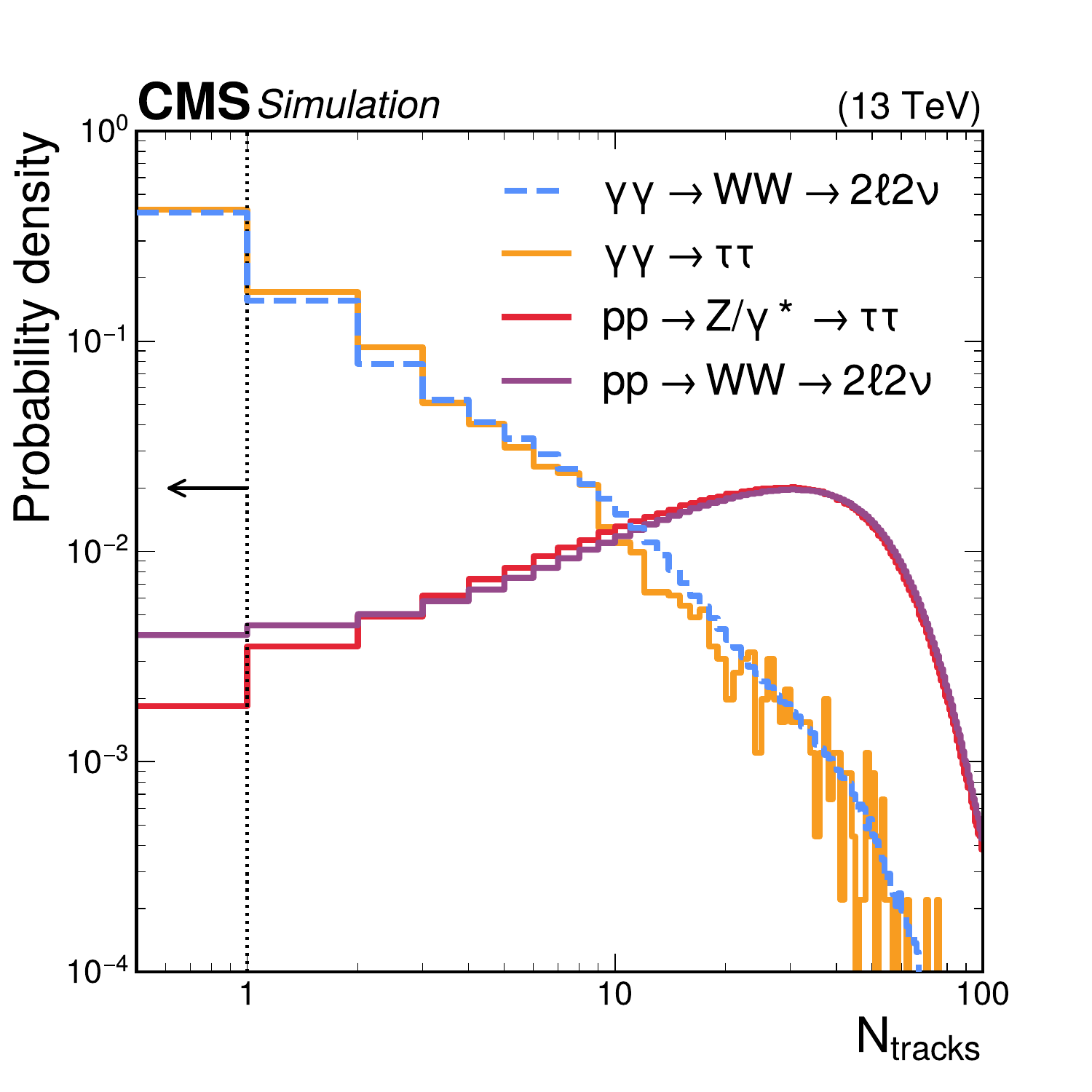}
  \includegraphics[width=0.45\textwidth]{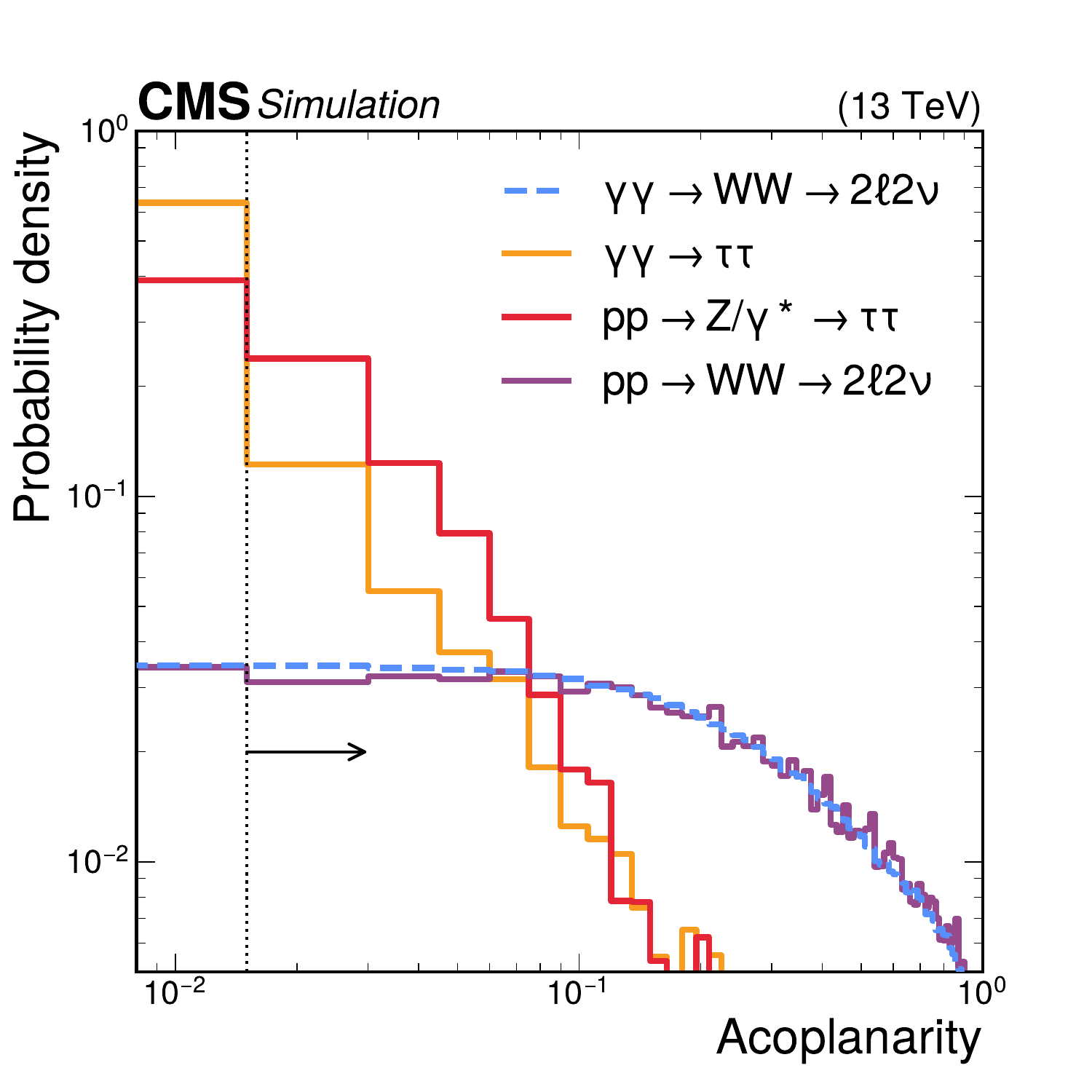}
	\caption{Normalized $\ntracks$ (left) and acoplanarity (right) distributions for simulated signal and background processes reconstructed in the detector with a PU distribution matching that of the data. The acoplanarity distribution is shown for events with $\ntracks=0$.\label{fig:sim}}
\end{figure}

The track multiplicity associated with the $\Pe\PGm$ vertex, $\ntracks$, is used to separate the signal from the backgrounds.
It is calculated considering tracks with $\abs{d_z}$ less than 0.05\cm with respect to the electron-muon vertex, $\pt>0.5\GeV$, and $\abs{\eta}<2.5$, and discarding the tracks used to build the selected electron and muon candidates. 
In the SR, $\ntracks$ is required to be zero. 
As shown in Fig.~\ref{fig:sim} (left), this requirement keeps about half of the elastic photon-fusion signal events because of pileup track 
contamination, while reducing quark- and gluon-induced inclusive processes by more than two orders of magnitude. 
The $\ntracks$ variable is poorly modelled by simulations, and data-driven corrections are determined and applied to signal and background samples, as detailed in Section \ref{sec:corr}. In the CR, $\ntracks$ is required to be between 1 and 5, which provides a CR enriched in inclusive backgrounds. An upper requirement of $\ntracks$ = 5 is imposed to select events with a similar topology as the SR and reduce extrapolation biases from the CR to the SR. The CR contains about 20 times more events than the SR.

In the $\ggtotautau$ process, the $\PGt$ leptons and their visible decay products are produced back-to-back~\cite{CMS:2024qjo}, while the decay products 
of the $\PW$ bosons from the $\ggtoww$ process are not. A minimum azimuth angle separation (acoplanarity) between the electron and the muon, 
$A = 1-\abs{\Delta\phi}/\pi>0.015$, is applied to reduce the $\ggtotautau$ contributions while keeping the signal efficiency higher than 
95\%, as shown in Fig.~\ref{fig:sim} (right). 
This selection criterion also removes a significant fraction of the DY background in the SR.

The $\PGm\PGm$ final state is used to derive corrections to the simulations based on data, as detailed in Section~\ref{sec:corr}. 
Events in the $\PGm\PGm$ final state are selected with single-muon triggers. Offline, the leading (subleading) muon must
have $\pt>26$--29 (10)\GeV depending on the data-taking year, and $\abs{\eta}<2.4$. 
The muons must have a transverse impact parameter with respect to the interaction vertex, $d_{xy}$, less than $0.05\cm$, 
and be separated from each other by at most 0.1\cm in the $z$
direction. The two muons must pass the identification and isolation criteria, and have OS charge and an invariant mass greater than 50\GeV. 

The event selection criteria for the $\Pe\PGm$ SR and CR, as well as for the $\PGm\PGm$ final state, are summarized in Table \ref{tab:sel}.

\begin{table}[ht]
	\topcaption{Selection criteria applied on the $\Pe\PGm$ SR [CR], and in the $\PGm\PGm$ final state. The \pt and pseudorapidity ranges correspond to different sets of triggers in different data-taking periods. The leading (subleading) lepton in the $\Pe\PGm$ SR [CR] must have $\pt>24$ (15)\GeV. }\label{tab:sel}
\centering
\begin{tabular}{llllll}
	& $\Pe\PGm$ SR [CR] & $\PGm\PGm$ \\
\hline
	$\pt^{\Pe}$ (\GeVns{}) & ${>}24(15)$  & \NA \\
$\abs{\eta^{\Pe}}$ & $<2.5$ & \NA \\
$\abs{d_{xy}^{\Pe}}$ (cm) & $<$0.05 in the barrel & \NA \\
 & $<$0.1 in the endcaps & \\
$\pt^{\PGm}$ (\GeVns{}) & ${>}24(15)$ & ${>}$26--29(10) \\
$\abs{\eta^{\PGm}}$ & $<$2.4 & $<$2.4 & \\
	$\abs{d_{xy}^{\PGm}}$ (cm) & $<$0.02 & $<$0.05\ \\
	$m_{\Pell\Pell}$ (\GeVns{}) & $>$20 & $>$50 \\
OS & Yes & Yes \\
$\abs{dz(\Pell,\Pell)}$ (cm) & $<$0.04  & $<$0.1 \\
$\Delta R$ & $>$0.5 & $>$0.5 \\
	$A$ & $>$0.015 & \NA \\
	$\ntracks$ & 0 [1--5] & \NA \\
\end{tabular}
\end{table}

\section{Analysis strategy}\label{sec:strategy}

The results of the analysis are extracted with a binned maximum likelihood fit of the $\ptemu$ distribution in the SR and the CR, 
where $\ptemu$ is defined as the magnitude of the vectorial $\pt$ sum of the selected electron and muon. 
The observable $\ptemu$ separates the processes with missing transverse momentum ($\ptmiss$), 
such as the signal and the inclusive $\PW\PW$ production from 
processes with low $\ptmiss$, such as the DY background, which dominate the low-$\ptemu$ bins, and the nonprompt background from multijet production.
Systematic uncertainties are relatively larger for $\ptmiss$ than $\ptemu$, which explains the choice of observable. 

In the maximum likelihood fits to extract the total and fiducial cross sections, the parameter of interest is the signal strength, while three other parameters are left free to float:
\begin{enumerate}
	\item the DY background normalization, correlated between the SR and the CR;
	\item the inclusive $\PW\PW$ normalization, correlated between the SR and the CR;
	\item the event weight correction for $\ntracks=0$ (described in Section~\ref{sec:corr}), correlated between the DY and the inclusive $\PW\PW$ backgrounds.
\end{enumerate}
The DY and inclusive $\PW\PW$ background normalizations are mostly constrained from the large CR sample, while the 
$\ntracks$ correction gets constrained by the DY background in the low-$\ptemu$ bins. 
In the likelihood scans to extract constraints on EFT operators, 
the SM $\ggtoww$ signal is constrained within uncertainties to its theoretical prediction and one EFT coefficient at a time, affecting the $\ggtoww$ process, is treated as a freely floating parameter of interest.

\section{Background and signal estimation}\label{sec:background}

\subsection{Corrections to the simulated track multiplicity}\label{sec:corr}

The pileup multiplicity in simulation is reweighted to match that observed in the data. After this correction, the 
simulations still do not describe accurately the $\ntracks$ distribution and need to be corrected further. 
Corrections to the multiplicity of tracks originating from the pileup ($\ntracksPU$) and from the hard scattering ($\ntracksHS$) vertices are 
derived as described in Ref.~\cite{CMS:2024qjo}, using events in the $\PGm\PGm$ final state.

The $\ntracksPU$ correction is obtained as an event weight 
by comparing the track multiplicity in data and in simulation inside 1-\mm wide windows 
sampled along the $z$ axis and at least 1\cm away from the dimuon vertex. It is applied to all simulated background and signal events.
The $\ntracksHS$ correction is applied as an event weight derived in a similar way, except the selected window is centered on the dimuon interaction vertex. 
This window includes both pileup and hard scattering tracks, and the pileup track multiplicity correction is applied first. This correction 
is measured for events with $A>0.015$, and applied to both DY and inclusive diboson events, which are modeled with the same generator settings. 
The $\ntracksHS$ correction reduces the event yields of the corresponding processes by about 40\% for events with $\ntracks=0$.

\subsection{Estimation of backgrounds with prompt leptons}

The DY background with prompt leptons is estimated from simulation, using the corrections described in Section \ref{sec:corr}. 
The inclusive quark-induced $\Pp\Pp\to\PW\PW$ ($2\Pell2\PGn$) background has a similar $\ptemu$ distribution as the signal, but, like the DY background, it is
strongly suppressed by the $\ntracks$ requirement.
The contribution from this process is estimated from simulation with the corrections described in Section \ref{sec:corr}. 
The same $\ntracksHS$ correction as that measured for the DY process is used, since the DY and diboson backgrounds are both simulated with 
the same generators and settings, and their $\pt(\PGt\PGt/\PW\PW)$ distributions are similar after selection. 
To make the analysis less dependent on the correction to $\ntracksHS$ and its uncertainty, which is 
difficult to evaluate because of the generator dependence, the $\ntracksHS$ correction to events with $\ntracks=0$ is left as a free parameter  
in the maximum likelihood fit described in Section \ref{sec:strategy}. 

Other backgrounds, such as $\ggtotautau$,  minor diboson processes ($\PW\PZ$, $\PZ\PZ$), single top quark, and $\ttbar$ production, 
are estimated from simulation, with corrections to the simulated $\ntracks$. For the $\ggtotautau$ process, which was measured directly in Ref.~\cite{CMS:2024qjo}, the simulated prediction (\textsc{SuperChic}~4) reproduces the measured cross section and kinematic distributions within the experimental and modeling uncertainties. Corrections to $\ntracksHS$ are not used for the single top quark and $\ttbar$ processes because these processes do not use the same generator. They are assigned large uncertainties (50\%) to cover for $\ntracksHS$ 
mismodelings, which have a negligible 
effect on the total background prediction. 

\subsection{Estimation of backgrounds with nonprompt leptons}

The background with jets misidentified as electrons or muons, essentially coming from $\PW + \text{jets}$ and QCD multijet events, is estimated from observed events.
Events where the electron and muon have same-sign (SS) electric charge are reweighted with an OS-to-SS scale factor (SF), denoted OSSS(id. $\Pe$, iso. $\PGm$).
Most events in the SS region correspond to the jet misidentification (mis-ID) background, but small contributions from other processes with genuine leptons
are estimated from simulation and subtracted from the data.

The OS-to-SS SF depends on both the muon and electron \pt and is measured in a background-enriched region 
where the muon relative isolation is
inverted (between 0.2 and 0.5 instead of ${<}0.2$ in the SR and CR) as:
\begin{equation}
	\text{OSSS (id. }\Pe\text{, anti-iso. }\PGm) = \frac{N^\text{OS}\text{(id. }\Pe\text{, anti-iso. }\PGm)}{N^\textrm{SS}\textrm{(id. }\Pe\text{, anti-iso. }\PGm)}, 
\end{equation}
where $N^\text{OS}\text{(id. }\Pe\text{, anti-iso. }\PGm)$ and 
$N^\text{SS}\text{(id. }\Pe\text{, anti-iso. }\PGm)$ are the number of OS and SS events with a well-identified electron and an 
anti-isolated muon. 
Inverting the muon isolation biases the OS-to-SS SFs, and a multiplicative correction to this SF is derived using events 
with an inverted electron identification. The OSSS(id. $\Pe$, iso. $\PGm$) term is computed as: 
\begin{equation}
	\text{OSSS (id. }\Pe\text{, iso. }\PGm) = \text{OSSS (id. }\Pe\textrm{, anti-iso. }\PGm) \frac{\text{OSSS (anti-id. }\Pe\text{, iso.}\PGm)}{\text{OSSS (anti-id. }\Pe\text{, anti-iso. }\PGm)}.
\end{equation}
In this expression, all the OSSS terms depend on the electron and muon \pt. 
They were verified not to depend on $\ntracks$. 

To improve the statistical precision of the mis-ID background distribution, its $\ptemu$ distribution is taken from events with $\ntracks<10$,
while the normalization is taken from the default estimation without relaxing the $\ntracks$ selection. 
The compatibility of the $\ptemu$ distributions in the SR with those in the region where events have $\ntracks<10$ was confirmed.

\section{Systematic uncertainties}\label{sec:systematics}

Systematic uncertainties are considered as nuisance parameters in the statistical procedure to extract the signal cross section. 
They are treated with either Gaussian (shape uncertainties) or log-normal (normalization uncertainties)
function priors included in the likelihood function~\cite{Cowan:2010js}. 
Unless specified, systematic uncertainties are treated as shape uncertainties affecting the $\ptemu$ distributions. Given that the cross section normalizations of the DY and diboson processes are both left floating in the 
maximum likelihood fit, overall normalization uncertainties are not considered for these processes. The sources of systematic uncertainties 
are summarized in Table~\ref{tab:sys}.

\begin{table}
\topcaption{Systematic uncertainties and their relative effects on the inclusive cross section measurement. An uncertainty of 2\% in the signal cross section is considered in the EFT interpretation.}
\label{tab:sys}
\centering
\begin{tabular}{lc}
Sources of uncertainties & Cross section uncertainty (\%) \\
\hline
\noalign{\vskip 1mm}
$\ntracksHS$ correction & 3.6 \\
DY normalization & 3.5 \\
$\Pe$ selection & 2.2 \\
$\ntracksPU$ correction & 2.0 \\
Integrated luminosity & 1.6 \\
$\ttbar$ normalization & 1.2 \\
Pileup modeling & 1.2 \\
Inclusive diboson normalization & 1.0 \\
$\PGm$ selection & 1.0 \\
Jet mis-ID prediction & $<$1 \\
$\mu_\mathrm{R}$, $\mu_\mathrm{F}$, PDF & $<$1 \\
\end{tabular}
\end{table}

The uncertainties in the electron and muon identification,
isolation, and triggering efficiencies are up to 2\% per lepton. 
A multiplicative correction to the identification and isolation SFs to account for the low-$\ntracks$ environment is applied to 
electrons; its magnitude is 0.98. It is derived as the ratio of the SFs measured in $\PZ\to\Pe\Pe$ events with $\ntracks<10$ and inclusively in $\ntracks$.  Scale factors for muons do not need to be
adjusted for the low-$\ntracks$ environment. 

The uncertainty in the $\ntracksPU$ correction amounts to 2\%, whereas no uncertainty in the $\ntracksHS$ correction is provided since it 
is freely floating in the SR with $\ntracks=0$. 
Theoretical uncertainties arising from the choice of renormalization and factorization scales ($\mu_\mathrm{R}$ and $\mu_\mathrm{F}$) used in the calculations, as well as from the PDF choice, are also taken into account and propagated to the $\ptemu$ distributions. 

Statistical uncertainties in the two-dimensional \pt-dependent OS-to-SS scaling factors used to evaluate the jet mis-ID background are included. An 
additional 10\% yield uncertainty is included to cover for a potential difference in the background composition in the SR and CR. It is determined by comparing 
OS-to-SS scaling factors measured in events with high or low $\ptmiss$.

The uncertainty in the cross section of the photon-fusion processes modeled with the \textsc{SuperChic}~4 generator is obtained by varying 
the rescattering models of the proton survival factors. Following the recommendations of Ref.~\cite{Shao:2022cly}, conservative uncertainties of 1\% for the elastic component, 10\% for semi-dissociative, and 50\% for fully dissociative production are assigned. Since the elastic, semi-dissociative, and fully dissociative components have similar $\ptemu$ distributions, these cross section uncertainties are applied only in the determination of the constraints on EFT operators, 
presented in Section \ref{sec:eftresults}, and not in the cross section measurements.

Statistical uncertainties in the number of simulated background events selected in the SR and CR or observed event yields in the regions used to derive the jet mis-ID background
are considered in all bins of the distributions using the approach from Ref.~\cite{BARLOW1993219,Conway:2011in}.
The systematic uncertainty in the pileup modeling is included by varying by $\pm$4.6\% the total inelastic cross section used to calculate simulated pileup
distributions~\cite{Sirunyan:2018nqx}.
The integrated luminosities for the 2016, 2017, and 2018 data-taking years have 1.2--2.5\% individual uncertainties~\cite{CMS-LUM-17-003,CMS-PAS-LUM-17-004,CMS-PAS-LUM-18-002}, whereas the overall uncertainty is 1.6\%.

\section{Results}\label{sec:results}

\subsection{Cross section measurement}

\begin{figure}
  \centering
  \includegraphics[width=0.45\textwidth]{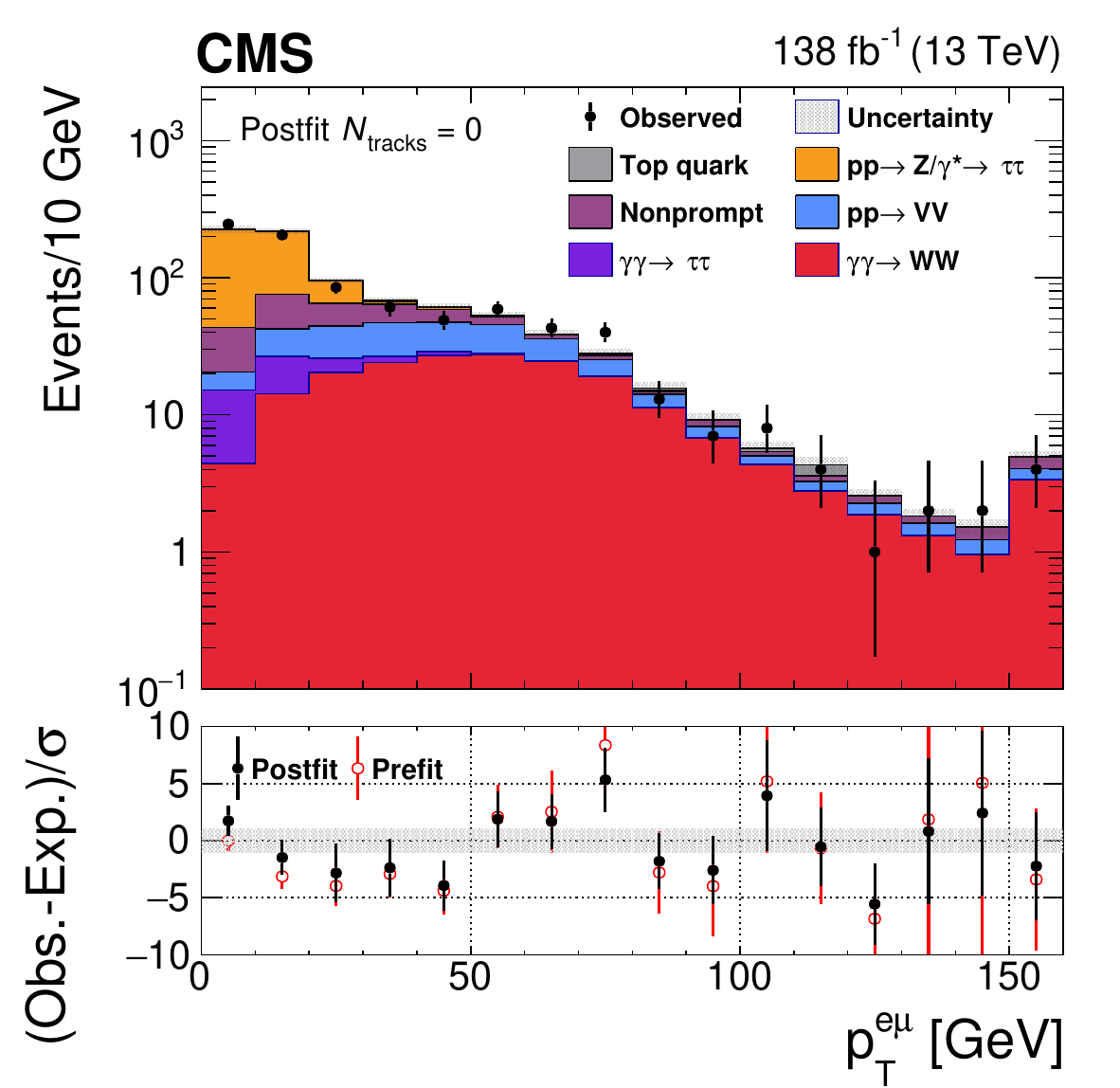}
  \includegraphics[width=0.45\textwidth]{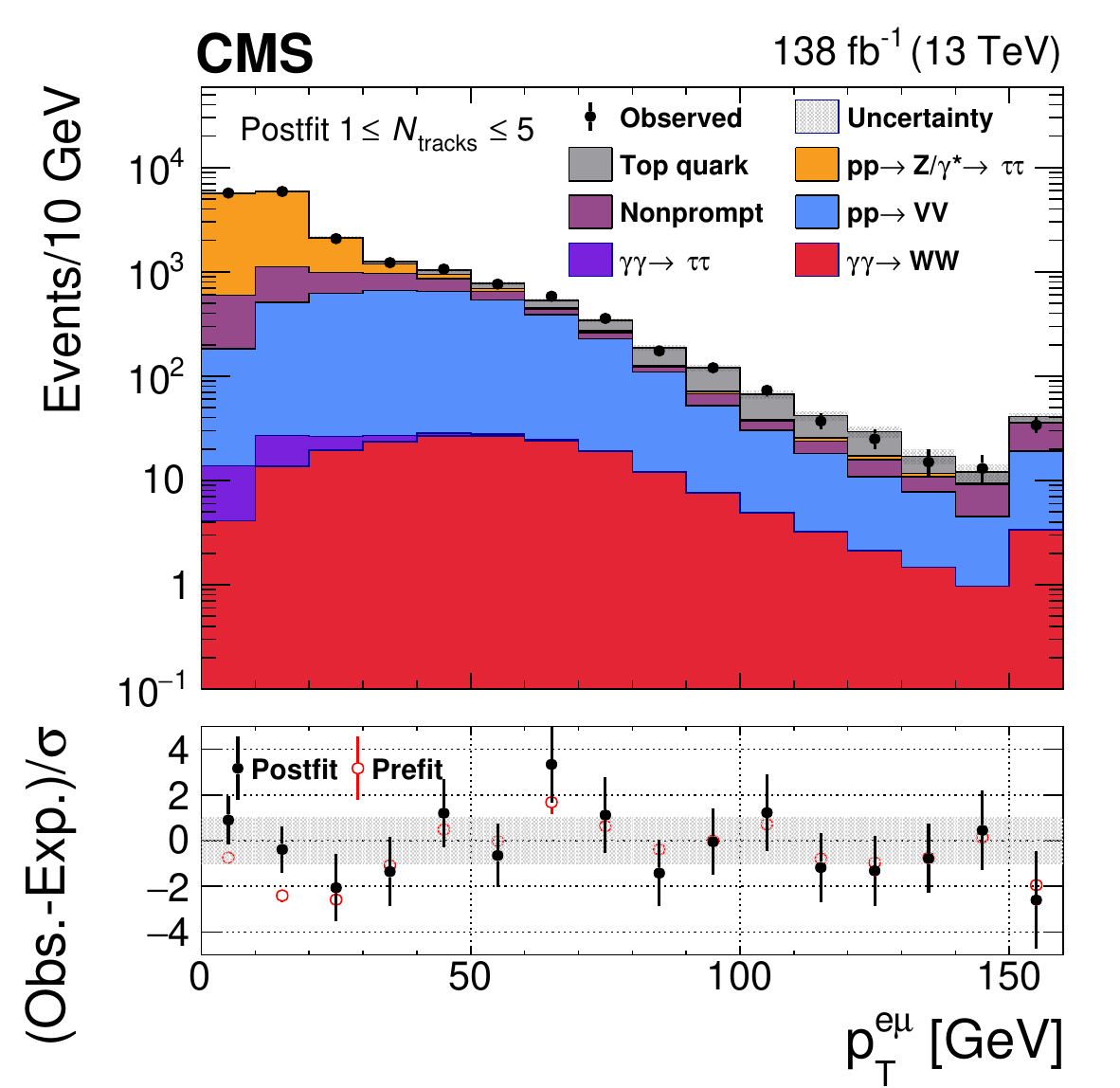}
	\caption{Observed and predicted $\ptemu$ distributions for events with $\ntracks=0$ (SR, left) or $1\leq\ntracks\leq5$ (CR, right), using 2016--2018 data. The distributions are shown after the maximum likelihood fit to the data ("postfit distributions"). The observed data and their associated Poissonian statistical uncertainty are shown with black markers with vertical error bars. The uncertainty band accouts for all sources of background and signal uncertainty, systematic as well as statistical, after the fit. The last bin includes the overflow. The lower panels show the ratio of data to sum of signal and background contributions, before (prefit, open red circles) and after (black full markers) the maximum likelihood fit.\label{fig:results_run2}}
\end{figure}

The inclusive and fiducial cross sections of the $\ggtoww$ process are extracted with the binned maximum likelihood fit described in Section \ref{sec:strategy}, using the $\ptemu$ distributions in the SR and CR shown in Fig.~\ref{fig:results_run2}. In the maximum likelihood fit, the 
\ntracksHS correction is pulled down to $0.91\pm 0.06$ times its initial value of $0.60$. The experimental data agree well with the sum of signal and background contributions for all $\ptemu$ regions. In the SR, the signal contribution dominates for $\ptemu>80\GeV$.  
The results presented hereafter have been determined from the measured $\ptemu$ distributions using the CMS statistical analysis tool \textsc{Combine}~\cite{CMS:2024onh}, 
which is based on the \textsc{RooFit}~\cite{Verkerke:2003ir} and \textsc{RooStats}~\cite{Moneta:2010pm} frameworks.
The sources of systematic uncertainties described in Section~\ref{sec:systematics} are included in the statistical procedure as nuisance parameters.

\begin{figure}
  \centering
  \includegraphics[width=0.65\textwidth]{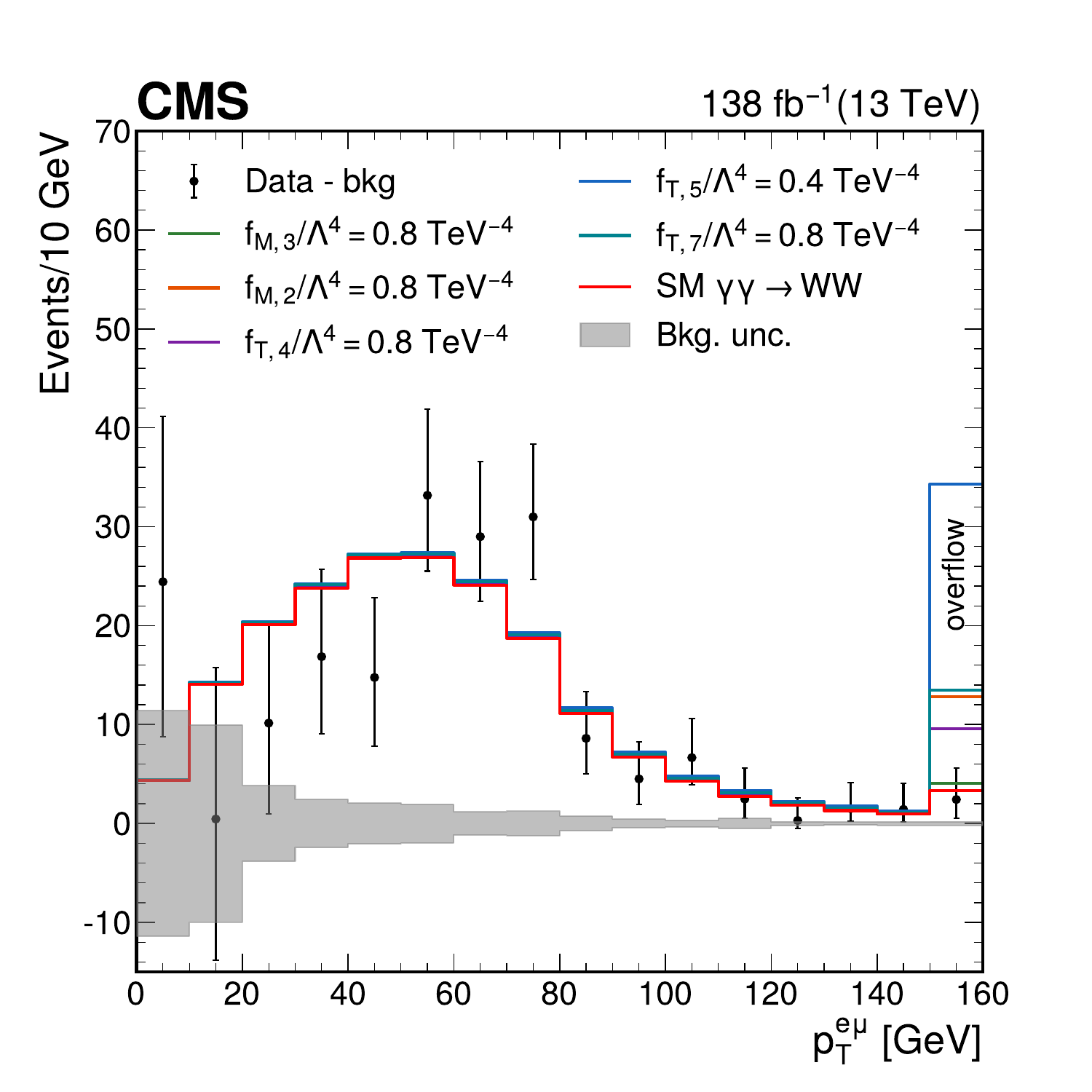}
        \caption{Distribution of $\ptemu$ for the difference between
	the observed number of events and the expected number of background events (black markers) compared with the SM prefit $\ggtoww$ prediction (red line), and $\ggtoww$ predictions for five benchmark points with nonzero EFT coefficients. The Poissonian statistical uncertainties associated with the observed data are represented by vertical error bars. The difference between the predictions for the SM and nonzero EFT coefficients is visible only in the overflow bin. The background uncertainty is shown with a gray band. \label{fig:eft}}
\end{figure}

The significance of the signal with respect to the background-only expectation is well above 5 standard deviations, and this process is observed for the first time using data collected with the CMS detector. The inclusive cross section for the $\ggtoww$ process is measured to be $643^{+82}_{-78}\unit{fb}$, 
whereas the value predicted by the \textsc{SuperChic}~4 generator is $631\pm 126\unit{fb}$~\cite{Bailey:2022wqy}. 
Figure~\ref{fig:eft} 
shows how the predicted $\ggtoww$ signal matches the difference between the observed events and
the simulated background events in the $\ptemu$ distributions.

The cross section of the $\ggtoww$ process is also measured in a fiducial phase space matching the analysis selection criteria
to the reconstructed event
selection. All variables used in the definition of the fiducial region are calculated
at the generator level after parton showering and hadronization, and the lepton momentum
includes the momenta of photons radiated within a cone of $\Delta R < 0.1$ centered on the lepton.
The fiducial region comprises events where the two $\PW$ bosons decay leptonically in the final state of an electron and a muon of OS charge. Events in which the electron or muon is produced via a leptonic $\PGt$ decay with the $\PGt$ lepton originating from a $\PW$ boson decay 
are also part of the signal 
within the fiducial region. 
The electron and muon at the generator level are required to have an acoplanarity $A>0.015$ and
an invariant mass ${>}20\GeV$.
There must not be any final-state charged particle with $c\tau>1\unit{cm}$, $\pt>0.5\GeV$, and $\abs{\eta}<2.5$, 
other than the two charged leptons themselves.
The leading lepton must satisfy $\pt>24\GeV$, whereas the requirement for the subleading lepton is $\pt>15\GeV$. The electron (muon) must have $\abs{\eta}<2.5$ (2.4).
Photon-induced diboson events not belonging to the fiducial region but entering the reconstruction-level categories are constrained to their expected normalizations and distributions, within uncertainties.
In the SR, the nonfiducial fraction of the $\ggtoww$
process is about 6\%.
The measured fiducial cross section is $\sigma^\text{fid}=3.96^{+0.53}_{-0.51}\unit{fb}$, whereas the value predicted by the \textsc{SuperChic}~4 generator is $3.87\pm 0.77\unit{fb}$~\cite{Bailey:2022wqy}.
The event yields after the maximum likelihood fit are shown in Table~\ref{tab:yields}.  

\begin{table}
	\topcaption{Observed and predicted event yields in the SR and CR. The signal and background yields are the result of the
global fit including all sources of uncertainties. }\label{tab:yields}
\centering
\begin{tabular}{lcc}
	Process & $\ntracks=0$ (SR) & $1\leq\ntracks\leq5$ (CR)    \\
	\hline
	
	 $\PZ/\PGgst\to\PGt\PGt$ & $361 \pm 22$ & $11\,260 \pm 170$ \\
	
	 Jet mis-ID & $119 \pm 10$ & $2\,160 \pm 180$  \\
	
	 Inclusive VV & $117 \pm 10$ & $3\,770 \pm 220$  \\
	
	 $\PGg\PGg\to\PGt\PGt$ & $34.2 \pm 1.5$ & $37.3 \pm 1.5$  \\
	
	 Top & $6.6 \pm 1.4$ & $673 \pm 98$  \\
	
	 EW WW $+2$ jets & $3.6 \pm 0.7$ & $28.9 \pm 4.9$  \\
	 Nonfiducial $\PGg\PGg\to\PW\PW$ & $11.4 \pm 0.4$ & $70.4 \pm 2.4$  \\[\cmsTabSkip]
	
	 Total bkg. & $653 \pm 30$ & $18\,000 \pm 160$  \\[\cmsTabSkip]
	
     Fiducial $\PGg\PGg\to\PW\PW$ & $183 \pm 22$ & $121 \pm 15$  \\[\cmsTabSkip]
	 
	 \hline
	 
	 Total & $835 \pm 29$ & $18\,120 \pm 170$  \\
	
	Observed & 829 & 18\,112 \\

\end{tabular}
\end{table}

\subsection{Effective field theory interpretation}\label{sec:eftresults}

In the SM, the photon-fusion production of a pair of $\PW$ bosons arises from self-interactions of electroweak gauge bosons, 
in particular from the triple gauge coupling (TGC) $\PGg\PW\PW$ and the quartic gauge coupling (QGC) $\PGg\PGg\PW\PW$. 
Whereas the individual diagrams contributing to this process exhibit amplitudes that grow with energy, 
their combined SM contributions interfere destructively and preserve unitarity. 
This delicate balance between TGC and QGC terms can be easily disrupted by the presence of BSM physics.

Whereas constraints on $\PGg\PW\PW$ anomalous TGCs are typically derived from the inclusive production of $\PW\PGg$, $\PW\PW$, and 
electroweak production of $\PW+\text{jets}$~\cite{CMS:2021cxr,CMS:2020mxy,CMS:2019nep},
the $\ggtoww$ process is primarily sensitive to anomalous QGCs and thus probes various dimension-8 operators within the effective field theory framework.

The results are interpreted with an EFT approach as constraints on the dimension-8 operators described in Ref.~\cite{Durieux:2025zrg}. 
The selected operators are bi-quadratic in the Higgs and gauge field strengths (denoted $f_M$) or  
quartic in the gauge field strength (denoted $f_T$). The CP-odd operators are denoted $f_{\widetilde{\mathrm{M}}}$ and $f_{\widetilde{\mathrm{T}}}$. 
For CP-odd operators, the interference with the SM amplitude cancels since the analysis observables are not CP-sensitive. 
The results are therefore driven by the quadratic EFT contribution.

We use the elastic \textsc{gamma-UPC} simulations to compute, in each
\ptemu bin, how each EFT operator changes the signal prediction.
These relative shifts are applied as event weights to the elastic, semi-dissociative, and dissociative components of the signal modeled by the \textsc{SuperChic}~4 generator, assuming the identical EFT effects for the nonelastic components.
As a cross-check of this assumption, the EFT weights were compared using two different photon flux models: the elastic photon approximation 
and an inclusive photon PDF. Despite the different photon spectra predicted by these models, no significant difference in 
the EFT weights was observed as a function of $\ptemu$. It should be noted that, 
when going beyond the collinear photon approximation, 
additional diagrams contribute to the semi- and fully dissociative productions, which do not strictly correspond to the $\ggtoww$ 
topology, as discussed in Ref.~\cite{Bailey:2022wqy}. The quantitative impact of EFT corrections on these contributions 
has not been studied yet and, in the absence of such studies, the EFT effects are assumed in this analysis to be identical for the elastic 
and dissociative components.
The impact is largest at high \ptemu, as shown in Fig.~\ref{fig:eft}. 

\begin{figure}
  \centering
  \includegraphics[width=0.9\textwidth]{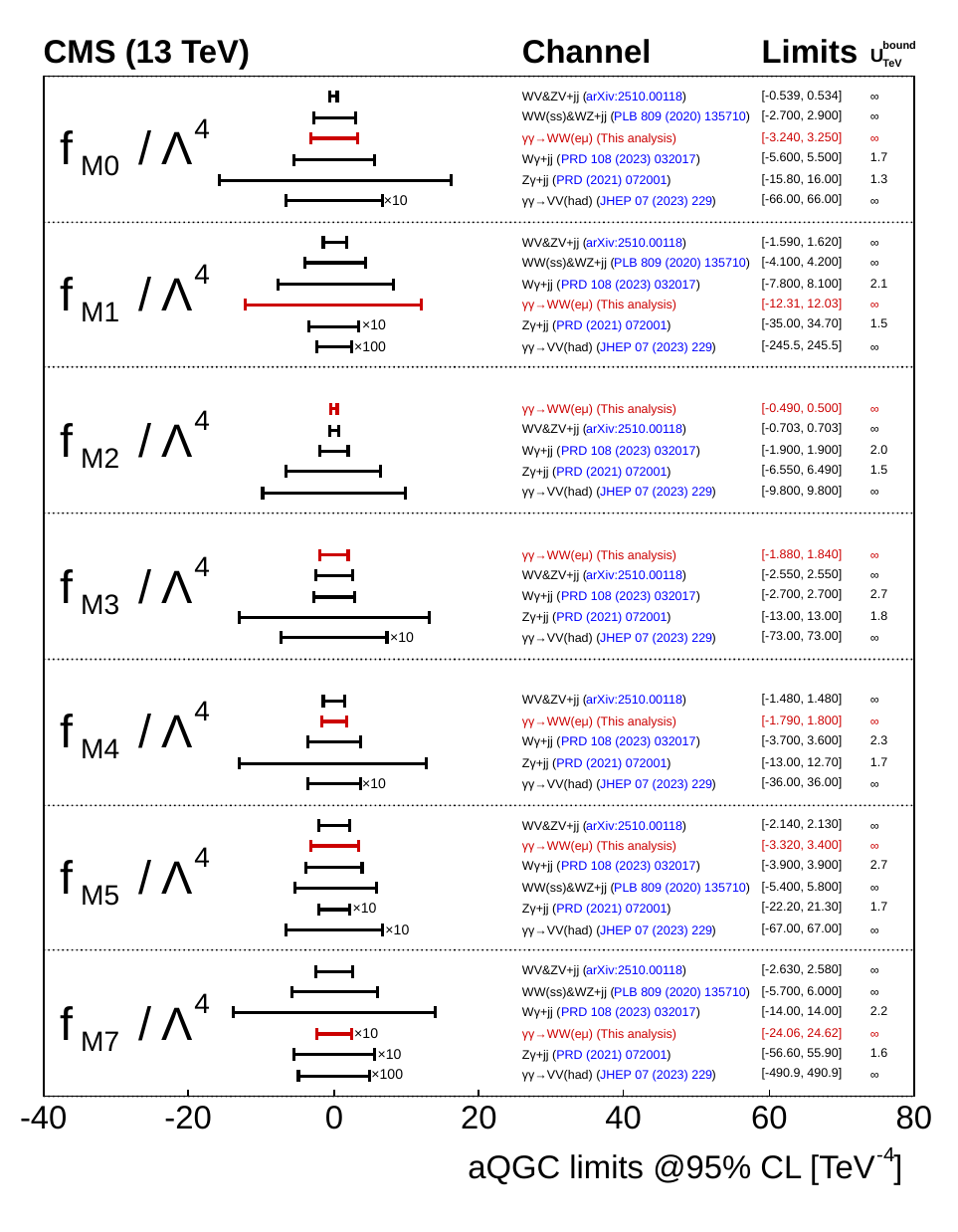}
	\caption{Summary of observed limits on the $f_\mathrm{M}$ Wilson coefficients for this analysis and previous results by the CMS Collaboration~\cite{CMS:2020gfh,CMS:2022yrl,CMS:2021gme,CMS:2022dmc,CMS:2025dbm}. The limits are arranged from most sensitive to least sensitive in descending order, with the current result highlighted in red. Horizontal error bars represent the 95\% confidence level intervals. In some cases, limits on the Wilson coefficients are shown under the assumption of unitarity bounds, as indicated in the right-most column. The unitarity bound is defined as the scattering energy at which the aQGC coupling strength is set equal to the observed limit that would result in a scattering amplitude that violates unitarity~\cite{Almeida_2020}. \label{fig:limits_compare_fM}}
\end{figure}

\begin{figure}
  \centering
  \includegraphics[width=0.9\textwidth]{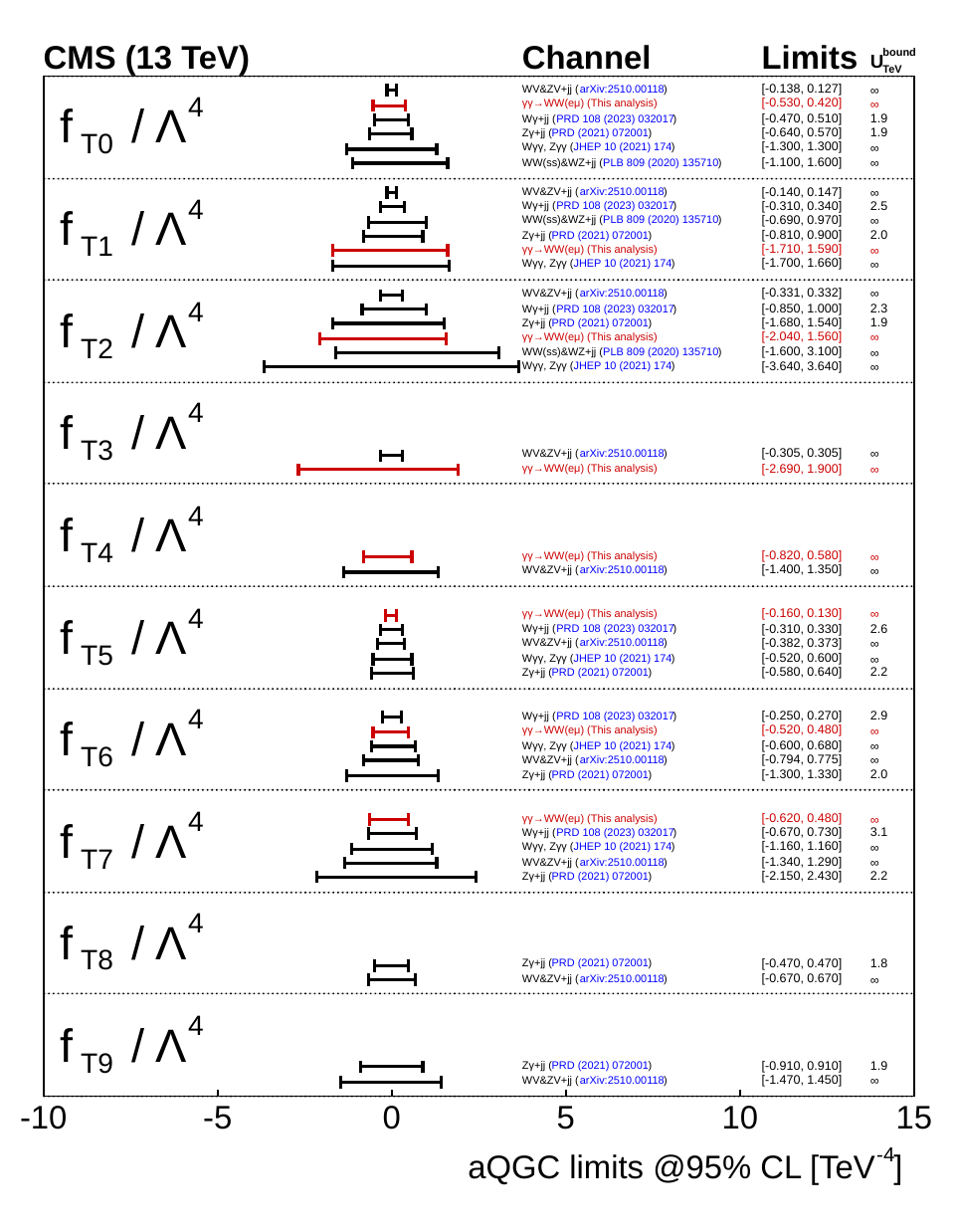}
        \caption{Summary of observed limits on the $f_\mathrm{T}$ Wilson coefficients for this analysis and previous results by the CMS Collaboration~\cite{CMS:2022yrl,CMS:2021gme,CMS:2021jji,CMS:2020gfh,CMS:2025dbm}. The limits are arranged from most sensitive to least sensitive in descending order, with the current result highlighted in red. Horizontal error bars represent the 95\% confidence level intervals. In some cases, limits on the Wilson coefficients are shown under the assumption of unitarity bounds, as indicated in the right-most column. The unitarity bound is defined as the scattering energy at which the aQGC coupling strength is set equal to the observed limit that would result in a scattering amplitude that violates unitarity~\cite{Almeida_2020}. \label{fig:limits_compare_fT}}
\end{figure}

The sensitivity of the measurement to the EFT operators, considered one at a time, 
is extracted from a likelihood scan performed by varying the corresponding EFT coefficient. 
The 95\% confidence-level (\CL) intervals on the EFT coefficients are extracted from the scan and reported
in Table~\ref{tab:eft}. Constraints on a series of operators are competitive with other measurements by the CMS Collaboration in different 
final states~\cite{CMS:2020gfh,CMS:2021jji,CMS:2021gme,CMS:2022yrl,CMS:2022dmc,CMS:2025dbm}. Figures~\ref{fig:limits_compare_fM} 
and ~\ref{fig:limits_compare_fT} show a comparison of the observed limits with those from other measurements.

\begin{table}
	\centering
	\topcaption{The 95\% \CL intervals on EFT operators varied one at a time while fixing the other ones to zero. Odd operators are denoted with a tilde sign.}\label{tab:eft}
	\begin{tabular}{lcc} & Expected 95\% \CL ($\TeVns^{-4}$) & Observed 95\% \CL ($\TeVns^{-4}$)\\
		\hline
	    $f_\mathrm{M0}/\Lambda^4$ & [$-4.03, 4.10$] & [$-3.24, 3.25$] \\
		$f_\mathrm{M1}/\Lambda^4$ & [$-15.46, 14.93$] & [$-12.31, 12.03$] \\
		$f_{\widetilde{\mathrm{M1}}}/\Lambda^4$ & [$-0.62, 0.62$] & [$-0.50, 0.49$] \\
		$f_\mathrm{M2}/\Lambda^4$ & [$-0.61, 0.63$] & [$-0.49, 0.50$] \\
		$f_{\widetilde{\mathrm{M2}}}/\Lambda^4$ & [$-2.24, 2.24$] & [$-1.80, 1.79$] \\
		$f_\mathrm{M3}/\Lambda^4$ & [$-2.36, 2.28$] & [$-1.88, 1.84$] \\
		$f_\mathrm{M4}/\Lambda^4$ & [$-2.22, 2.26$] & [$-1.79, 1.80$] \\
		$f_{\widetilde{\mathrm{M4}}}/\Lambda^4$ & [$-4.49, 4.49$] & [$-3.59, 3.58$] \\
		$f_\mathrm{M5}/\Lambda^4$ & [$-4.12, 4.27$] & [$-3.32, 3.40$] \\
		$f_{\widetilde{\mathrm{M5}}}/\Lambda^4$ & [$-2.03, 2.03$] & [$-1.63, 1.62$] \\
		$f_\mathrm{M7}/\Lambda^4$ & [$-29.86, 30.92$] & [$-24.06, 24.62$] \\
		$f_\mathrm{T0}/\Lambda^4$ & [$-0.65, 0.53$] & [$-0.53, 0.42$] \\
		$f_\mathrm{T1}/\Lambda^4$ & [$-2.13, 2.00$] & [$-1.71, 1.59$] \\
		$f_\mathrm{T2}/\Lambda^4$ & [$-2.50, 2.01$] & [$-2.04, 1.56$] \\
		$f_{\widetilde{\mathrm{T2}}}/\Lambda^4$ & [$-0.090, 0.090$] & [$-0.03, 0.03$] \\
		$f_\mathrm{T3}/\Lambda^4$ & [$-3.28, 2.49$] & [$-2.69, 1.90$] \\
		$f_{\widetilde{\mathrm{T3}}}/\Lambda^4$ & [$-0.089, 0.089$] & [$-0.071, 0.071$] \\
		$f_\mathrm{T4}/\Lambda^4$ & [$-1.00, 0.76$] & [$-0.82, 0.58$] \\
		$f_{\widetilde{\mathrm{T4}}}/\Lambda^4$ & [$-0.38, 0.38$] & [$-0.30, 0.30$] \\
		$f_\mathrm{T5}/\Lambda^4$ & [$-0.20, 0.16$] & [$-0.16, 0.13$] \\
		$f_{\widetilde{\mathrm{T5}}}/\Lambda^4$ & [$-0.21, 0.21$] & [$-0.17, 0.17$] \\
		$f_\mathrm{T6}/\Lambda^4$ & [$-0.65, 0.61$] & [$-0.52, 0.48$] \\
		$f_{\widetilde{\mathrm{T6}}}/\Lambda^4$ & [$-0.62, 0.62$] & [$-0.50, 0.50$] \\
		$f_\mathrm{T7}/\Lambda^4$ & [$-0.76, 0.61$] & [$-0.62, 0.47$] \\
        \end{tabular}
\end{table}

\section{Summary}~\label{sec:summary}

The observation of the photon-fusion production of $\PW$ boson pairs using the CMS detector at the LHC is reported. The measurement utilizes proton-proton collision data collected in 2016--2018, corresponding to an integrated luminosity of 138\fbinv at $\sqrt{s}=13\TeV$.
The experimental strategy exploits the exclusive nature of the signal, requiring the absence of additional charged particles 
associated with the dilepton production vertex. Events are selected in the final state with one isolated electron and one isolated muon, and no additional tracks associated with the electron-muon production vertex. 
The analysis employs a maximum likelihood fit to the transverse momentum sum of the electron and muon to extract the cross section 
of the signal.
The measured inclusive cross section for $\ggtoww$ production is $643^{+82}_{-78}\unit{fb}$, whereas the fiducial cross section, defined in a phase space matching the experimental selection criteria, is $3.96^{+0.53}_{-0.51}\unit{fb}$. These results show excellent agreement with the 
standard model predictions of $631\pm126\unit{fb}$ and $3.87\pm0.77\unit{fb}$ for the inclusive and fiducial cross sections, respectively.
Beyond the cross section measurement, this analysis provides the most stringent constraints to date on several 
anomalous quartic gauge couplings using dimension-8 operators in an effective field theory approach. 

\begin{acknowledgments}
We congratulate our colleagues in the CERN accelerator departments for the excellent performance of the LHC and thank the technical and administrative staffs at CERN and at other CMS institutes for their contributions to the success of the CMS effort. In addition, we gratefully acknowledge the computing centers and personnel of the Worldwide LHC Computing Grid and other centers for delivering so effectively the computing infrastructure essential to our analyses. Finally, we acknowledge the enduring support for the construction and operation of the LHC, the CMS detector, and the supporting computing infrastructure provided by the following funding agencies: SC (Armenia), BMBWF and FWF (Austria); FNRS and FWO (Belgium); CNPq, CAPES, FAPERJ, FAPERGS, and FAPESP (Brazil); MES and BNSF (Bulgaria); CERN; CAS, MoST, and NSFC (China); MINCIENCIAS (Colombia); MSES and CSF (Croatia); RIF (Cyprus); SENESCYT (Ecuador); ERC PRG, TARISTU24-TK10 and MoER TK202 (Estonia); Academy of Finland, MEC, and HIP (Finland); CEA and CNRS/IN2P3 (France); SRNSF (Georgia); BMFTR, DFG, and HGF (Germany); GSRI (Greece); NKFIH (Hungary); DAE and DST (India); IPM (Iran); SFI (Ireland); INFN (Italy); MSIT and NRF (Republic of Korea); MES (Latvia); LMTLT (Lithuania); MOE and UM (Malaysia); BUAP, CINVESTAV, CONACYT, LNS, SEP, and UASLP-FAI (Mexico); MOS (Montenegro); MBIE (New Zealand); PAEC (Pakistan); MES, NSC, and NAWA (Poland); FCT (Portugal);  MESTD (Serbia); MICIU/AEI and PCTI (Spain); MOSTR (Sri Lanka); Swiss Funding Agencies (Switzerland); MST (Taipei); MHESI (Thailand); TUBITAK and TENMAK (T\"{u}rkiye); NASU (Ukraine); STFC (United Kingdom); DOE and NSF (USA).

\hyphenation{Rachada-pisek} Individuals have received support from the Marie-Curie program and the European Research Council and Horizon 2020 Grant, contract Nos.\ 675440, 724704, 752730, 758316, 765710, 824093, 101115353, 101002207, 101001205, and COST Action CA16108 (European Union); the Leventis Foundation; the Alfred P.\ Sloan Foundation; the Alexander von Humboldt Foundation; the Science Committee, project no. 22rl-037 (Armenia); the Fonds pour la Formation \`a la Recherche dans l'Industrie et dans l'Agriculture (FRIA) and Fonds voor Wetenschappelijk Onderzoek contract No. 1228724N (Belgium); the Beijing Municipal Science \& Technology Commission, No. Z191100007219010, the Fundamental Research Funds for the Central Universities, the Ministry of Science and Technology of China under Grant No. 2023YFA1605804, the Natural Science Foundation of China under Grant No. 12535004, and USTC Research Funds of the Double First-Class Initiative No.\ YD2030002017 (China); the Ministry of Education, Youth and Sports (MEYS) of the Czech Republic; the Shota Rustaveli National Science Foundation, grant FR-22-985 (Georgia); the Deutsche Forschungsgemeinschaft (DFG), among others, under Germany's Excellence Strategy -- EXC 2121 ``Quantum Universe" -- 390833306, and under project number 400140256 - GRK2497; the Hellenic Foundation for Research and Innovation (HFRI), Project Number 2288 (Greece); the Hungarian Academy of Sciences, the New National Excellence Program - \'UNKP, the NKFIH research grants K 131991, K 133046, K 138136, K 143460, K 143477, K 146913, K 146914, K 147048, 2020-2.2.1-ED-2021-00181, TKP2021-NKTA-64, and 2025-1.1.5-NEMZ\_KI-2025-00004 (Hungary); the Council of Science and Industrial Research, India; ICSC -- National Research Center for High Performance Computing, Big Data and Quantum Computing, FAIR -- Future Artificial Intelligence Research, and CUP I53D23001070006 (Mission 4 Component 1), funded by the NextGenerationEU program (Italy); the Latvian Council of Science; the Ministry of Education and Science, project no. 2022/WK/14, and the National Science Center, contracts Opus 2021/41/B/ST2/01369, 2021/43/B/ST2/01552, 2023/49/B/ST2/03273, and the NAWA contract BPN/PPO/2021/1/00011 (Poland); the Funda\c{c}\~ao para a Ci\^encia e a Tecnologia, grant CEECIND/01334/2018 (Portugal); the National Priorities Research Program by Qatar National Research Fund; MICIU/AEI/10.13039/501100011033, ERDF/EU, ``European Union NextGenerationEU/PRTR", and Programa Severo Ochoa del Principado de Asturias (Spain); the Chulalongkorn Academic into Its 2nd Century Project Advancement Project, the National Science, Research and Innovation Fund program IND\_FF\_68\_369\_2300\_097, and the Program Management Unit for Human Resources \& Institutional Development, Research and Innovation, grant B39G680009 (Thailand); the Kavli Foundation; the Nvidia Corporation; the SuperMicro Corporation; the Welch Foundation, contract C-1845; and the Weston Havens Foundation (USA).
\end{acknowledgments}\section*{Data availability} Release and preservation of data used by the CMS Collaboration as the basis for publications is guided by the  \href{https://doi.org/10.7483/OPENDATA.CMS.1BNU.8V1W}{CMS data preservation, re-use and open access policy}.

\bibliography{auto_generated}

\cleardoublepage \appendix\section{The CMS Collaboration \label{app:collab}}\begin{sloppypar}\hyphenpenalty=5000\widowpenalty=500\clubpenalty=5000\cmsinstitute{Yerevan Physics Institute, Yerevan, Armenia}
{\tolerance=6000
A.~Hayrapetyan, V.~Makarenko\cmsorcid{0000-0002-8406-8605}, A.~Tumasyan\cmsAuthorMark{1}\cmsorcid{0009-0000-0684-6742}
\par}
\cmsinstitute{Institut f\"{u}r Hochenergiephysik, Vienna, Austria}
{\tolerance=6000
W.~Adam\cmsorcid{0000-0001-9099-4341}, L.~Benato\cmsorcid{0000-0001-5135-7489}, T.~Bergauer\cmsorcid{0000-0002-5786-0293}, M.~Dragicevic\cmsorcid{0000-0003-1967-6783}, C.~Giordano, P.S.~Hussain\cmsorcid{0000-0002-4825-5278}, M.~Jeitler\cmsAuthorMark{2}\cmsorcid{0000-0002-5141-9560}, N.~Krammer\cmsorcid{0000-0002-0548-0985}, A.~Li\cmsorcid{0000-0002-4547-116X}, D.~Liko\cmsorcid{0000-0002-3380-473X}, M.~Matthewman, J.~Schieck\cmsAuthorMark{2}\cmsorcid{0000-0002-1058-8093}, R.~Sch\"{o}fbeck\cmsAuthorMark{2}\cmsorcid{0000-0002-2332-8784}, M.~Shooshtari\cmsorcid{0009-0004-8882-4887}, M.~Sonawane\cmsorcid{0000-0003-0510-7010}, W.~Waltenberger\cmsorcid{0000-0002-6215-7228}, C.-E.~Wulz\cmsAuthorMark{2}\cmsorcid{0000-0001-9226-5812}
% ...
\par}
\vskip\cmsinstskip
\dag:~Deceased\\
$^{1}$Also at Yerevan State University, Yerevan, Armenia\\
$^{2}$Also at TU Wien, Vienna, Austria\\
% ...
\end{sloppypar}
\end{document}